\documentclass[aps,prd,twocolumn,superscriptaddress,preprintnumbers,floatfix,nofootinbib,notitlepage,showkeys,showpacs]{revtex4-1}

\usepackage[utf8x]{inputenc}

\usepackage{graphicx}
\usepackage{hyperref}
\usepackage{latexsym}
\usepackage{amsmath}
\usepackage{amssymb}
\usepackage{bbm}
\usepackage{ulem}
\usepackage{pdfsync}
\usepackage{epsfig}
\usepackage{epstopdf}
\usepackage{subfigure}
\usepackage{color}
\usepackage{comment}
\usepackage{slashed}

\newcommand{\be}{\begin{equation}}
\newcommand{\ee}{\end{equation}}
\newcommand{\bea}{\begin{equation}\begin{aligned}}
\newcommand{\eea}{\end{aligned}\end{equation}}

\def\lsim{\mathrel{\raise.3ex\hbox{$<$\kern-.75em\lower1ex\hbox{$\sim$}}}}
\def\gsim{\mathrel{\raise.3ex\hbox{$>$\kern-.75em\lower1ex\hbox{$\sim$}}}}
\begin{document}

\title{Reheating the Standard Model from a hidden sector}

\author{Tommi Tenkanen}
\email{tommi.tenkanen@helsinki.fi}
\affiliation{Department of Physics, University of Helsinki, \\ P.O.~Box 64, FI-00014, Helsinki, Finland}
\affiliation{Helsinki Institute of Physics, \\ P.O.~Box 64, FI-00014, Helsinki, Finland}   
\author{Ville Vaskonen}
\email{ville.vaskonen@jyu.fi}
\affiliation{Helsinki Institute of Physics, \\ P.O.~Box 64, FI-00014, Helsinki, Finland}  
\affiliation{University of Jyvaskyla, Department of Physics, \\ P.O.~Box 35 (YFL), FI-40014  Jyv\"askyl\"a, Finland}

\begin{abstract}
We consider a scenario where the inflaton decays to a hidden sector thermally decoupled from the visible Standard Model sector. A tiny portal coupling between the hidden and the visible sectors later heats the visible sector so that the Standard Model degrees of freedom come to dominate the energy density of the Universe before Big Bang Nucleosynthesis. We find that this scenario is viable, although obtaining the correct dark matter abundance and retaining successful Big Bang Nucleosynthesis is not obvious. We also show that the isocurvature perturbations constituted by a primordial Higgs condensate are not problematic for the viability of the scenario.
\end{abstract}

\preprint{HIP-2016-18/TH}

\maketitle


\section{Introduction}

Despite of large observational evidence for the existence of dark matter (DM) and for the occurrence of cosmic inflation in the past, their nature and properties remain to a large extent unknown. While the inflationary dynamics can be successfully explained within the Standard Model (SM) of particle physics \cite{Bezrukov:2007ep}, the large non-baryonic matter content of our Universe, i.e. dark matter, can not (see e.g. \cite{Klasen:2015uma} for a recent review). This requires one to extend the SM by e.g. assuming a hidden sector which might have played an important role in the early Universe but which has thus far evaded all current observational bounds. 

In principle, there is no reason to expect detecting such a sector by current experiments. While for example the $750$\,GeV diphoton excess at the LHC \cite{ATLAS-CONF-2015-081,CMS-PAS-EXO-15-004} was encouraging, new physics might remain undetected by all present and future colliders, or new particles might belong to a sector which comprises only a part of the correct extension of the SM.

Depending on interactions between the SM and new physics, new sectors can provide interesting alternatives for the thermal history of our Universe. By the standard lore, soon after the cosmic inflation the SM sector forms an equilibrium heat bath during a process called reheating, where the inflaton field decays to SM particles either directly or via mediator fields \cite{Kofman:1994rk,Kofman:1997yn}. For a successful Big Bang Nucleosynthesis (BBN) the SM has to become the dominant energy density component before $T_{\rm SM}\simeq 4$\,MeV \cite{Kawasaki:2000en,Hannestad:2004px,Ichikawa:2005vw,DeBernardis:2008zz}.

Usually, all decay products including possible dark matter candidates are assumed to become part of the same heat bath. The observed dark matter abundance is then obtained via the freeze-out mechanism, where dark matter particles decouple from the SM bath when their mutual interaction rate can not compete with the expansion rate of the Universe any more. While alternatives for this scenario exist, such as the freeze-in production of dark matter \cite{McDonald:2001vt, Hall:2009bx} or asymmetric reheating \cite{Berezhiani:1995am,Feng:2008mu,Adshead:2016xxj}, it has become customary to assume that the SM reaches thermal equilibrium at a relatively early stage and governs the evolution of the Universe from that point on. A hidden sector thermally decoupled from the SM sector is, however, as plausible a candidate for explaining the whole thermal history of the Universe down to BBN, including cosmic inflation, reheating, and production of dark matter.

In this work, we consider a scenario where the inflaton decays to a hidden sector ultraweakly coupled to the visible sector. A similar scenario was recently studied in \citep{Berlin:2016vnh}, where the primary goal was to show that even as massive as PeV-scale particles can comprise the observed dark matter relic density. We confront this scenario against observational constraints including not only dark matter abundance and BBN, but also isocurvature perturbations constituted by a primordial Higgs condensate. Studying the effect of the Higgs condensate is crucial in determining the viability of this scenario as scalar fields are generically known to acquire large vacuum expectation values during cosmic inflation \cite{Starobinsky:1994bd}, possibly enabling large isocurvature modes between DM and baryon-photon fluid. We find that a hidden sector thermally decoupled from the SM sector is in good agreement with all the above constraints.

The paper is organized as follows: in Sec. \ref{reheating} we set up the scenario and discuss how the SM sector was reheated and the observed DM abundance produced in the early Universe. In Sec. \ref{isocSection} we confront this scenario against the observational bound for primordial isocurvature perturbations by considering the inflationary behaviour of the Higgs field. Finally, we present our results in Sec. \ref{resultsSection} and conclusions in Sec. \ref{conclusions}.

\section{Reheating the Standard Model}
\label{reheating}

We consider a scenario in which the decay of the inflaton field populates the hidden sector soon after the cosmic inflation. We assume the inflaton decays to the visible SM sector only in negligible amounts so that the SM sector remains energetically subdominant. The hidden sector very rapidly thermalizes to a temperature $T_{\rm h}$ and starts to govern evolution of the Universe. As the Universe expands, scattering rates within the hidden sector eventually become smaller than the Hubble rate. At some point dark matter freezes out in the hidden sector, and decays and annihilations of hidden sector particles heat up the SM sector. After this, the standard Hot Big Bang scenario is recovered.

For concreteness, we consider a class of models where the hidden sector couples to the SM only via Higgs portal. In this section, we discuss the SM heating and DM production in two parts: first, by considering the simplest possible scenario where the hidden sector consists of a $Z_2$ symmetric scalar only, and then by allowing for a general renormalizable scalar potential and a fermionic DM candidate in the hidden sector.

\subsection{Scalar dark matter}
\label{scalarcase}

Let us consider the simplest possible model where the hidden sector consists only of a $Z_2$ symmetric scalar field $s$, which couples to the SM Higgs field $h$ via the portal $\lambda_{\rm hs} s^2 h^2$. The annihilations of $s$ to SM particles then heat up the SM sector. The heating stops as the hidden sector temperature drops below the mass of the final state particles (or mass of $s$ if it is heavier than the produced particles), and finally as the rate of the $ssss\to ss$ process becomes smaller than the Hubble rate, the $s$ abundance freezes out to comprise the dark matter abundance we observe today. This kind of dark matter freeze-out via number-changing self-interactions has been studied in e.g \cite{Chu:2011be,Bernal:2015bla,Bernal:2015ova,Bernal:2015xba,Heikinheimo:2016yds}.

To characterize the evolution of energy densities of the hidden and visible sectors, we solve the Boltzmann equations for the $s$ number density $n_{\rm s}$, and SM energy density 
\be \label{rhogamma}
\rho_\gamma = \frac{\pi^2}{30} g_* T_{\rm SM}^4,
\ee 
where $g_*$ counts for the effective number of relativistic degrees of freedom in the SM sector at temperature $T_{\rm SM}$. The SM energy density can be solved from \cite{Gondolo:1990dk}
\be
\dot\rho_\gamma + 4H\rho_\gamma = \frac{T_{\rm h}}{64\pi^4} \int\mathrm{d}s' s'(s'-4m_{\rm s}^2) \sigma_s K_1(\sqrt{s'}/T_{\rm h}) ,
\ee
assuming that $s$ remains thermal. Here $\sigma_s=\sigma_{ss\to hh} + \sigma_{ss\to WW} + \sigma_{ss\to ZZ} + \sigma_{ss\to bb} + \dots$ is the $s$ annihilation cross section \cite{Cline:2012hg}, $H^2 = 8\pi(\rho_\gamma+\rho_{\rm s})/(3M_{\rm P}^2)$, and $M_{\rm P}\approx 1.22\times10^{19}$\,GeV is the Planck mass. 

\begin{figure}
\begin{center}
\includegraphics[width = 0.46\textwidth]{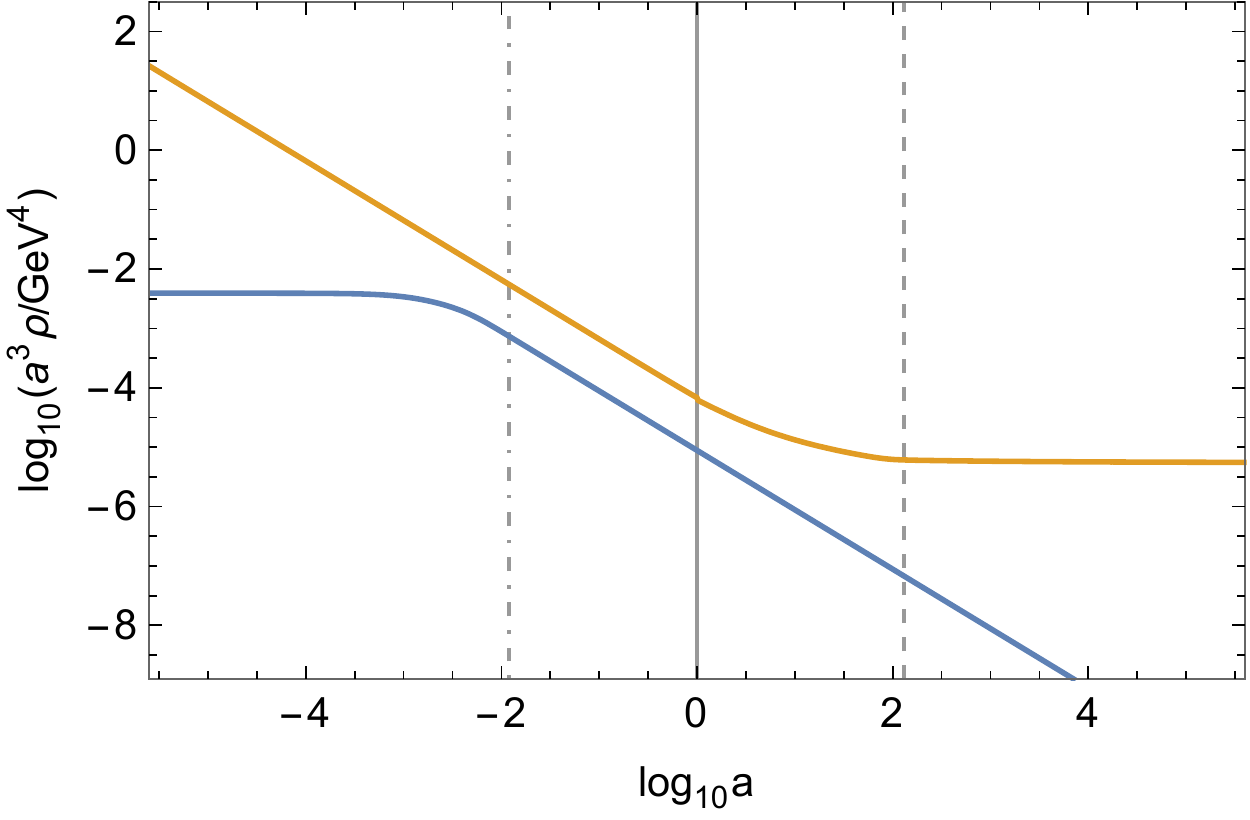}
\caption{Blue and yellow lines show the evolution of the SM and $Z_2$-symmetric $s$ energy densities, respectively. The dot-dashed line shows when $T_{\rm h}=2m_{\rm b}$, the solid line shows when $s$ becomes non-relativistic, and the dashed line marks the freeze-out of $s$ at $\tfrac{1}{4!}n^3_{\rm s}\left\langle v^3\sigma_{ssss\to ss}\right\rangle=H$. Here $m_{\rm s} = 0.1$\,GeV, $\lambda_{\rm hs} = 10^{-7}$, and $\lambda_{\rm s} = 1$.}
\label{freezeoutZ2}
\end{center}
\end{figure}

Similar to Ref. \cite{Carlson:1992fn}, we relate the hidden sector temperature $T_{\rm h}$ to the scale factor $a$ via conservation of entropy, $\tilde s a^3 = $ constant. Note that we need the hidden sector temperature only to calculate the equilibrium distribution of $s$ particles. Hence, for the entropy density, $\tilde s(T_{\rm h})$, we can use the usual equilibrium relation. We normalize $a$ such that $a = 1$ when $T_{\rm h}=m_{\rm s}$. Before $s$ becomes non-relativistic the hidden sector temperature scales as $T_{\rm h} = m_{\rm s}/a$, but after that $T_{\rm h}$ drops only logarithmically as a function of $a$ until $s$ freezes out. We solve the $s$ number density from
\be
\dot{n}_{\rm s} + 3Hn_{\rm s} = -\tfrac{1}{4!} \left\langle v^3\sigma_{ssss\to ss}\right\rangle (n_{\rm s}^4 - n_{\rm s}^2(n_{\rm s}^{\rm eq})^2) ,
\ee
where $n_{\rm s}^{\rm eq}$ is the equilibrium number density which we evaluate numerically, and where in the non-relativistic limit $\left\langle v^3\sigma_{ssss\to ss}\right\rangle \simeq \lambda_{\rm s}^4/m_{\rm s}^8$. The generic evolution of the SM and hidden sector energy densities is depicted in Fig. \ref{freezeoutZ2}. 

The energy density of $s$ particles scales as radiation, $a^{-4}$, until the temperature of the $s$ bath drops below $m_{\rm s}$. Then, the $s$ energy density scales as $1/(a^{3}\ln a)$ until the $s$ freeze-out, after which it scales as cold dark matter, $a^{-3}$. The SM energy density produced via $s$ annihilations is approximately (see e.g. \cite{Chu:2011be})
\be \label{sannihilations}
\rho_\gamma \simeq \rho_{\rm s} \frac{n_{\rm s} \left\langle v\sigma_s\right\rangle}{H}, 
\ee
Because of particle production from $s$ annihilations, the SM energy density first scales as $a^{-3}$, as the $s$ annihilation cross section to SM particles, $\sigma_s$, scales as $1/s'\sim a^2$ for large $T_{\rm h}$. After the end of $s$ annihilations $\rho_\gamma$ scales as $a^{-4}$.

To accomplish successful nucleosynthesis and to recover the observed peak structure of the Cosmic Microwave Background radiation, the SM energy density should dominate the energy density of the Universe from BBN at $T_{\rm SM}\sim 4$\,MeV to the matter-radiation equality at $T_{\rm SM}\sim 0.8$ eV. From Eq. \eqref{sannihilations} it is obvious, that obtaining $\rho_\gamma>\rho_{\rm s}$ by $s$ annihilations requires $n_{\rm s} \left\langle v\sigma_s\right\rangle\gtrsim H$. Hence, the only way to get this scenario work is by having a large enough portal coupling $\lambda_{\rm hs}$ to thermalize the SM sector with the $s$ bath. In that case, we end up in the standard DM freeze-out scenario studied extensively in the case of a $Z_2$ symmetric scalar field \cite{McDonald:1993ex,Burgess:2000yq,McDonald:2001vt,Cline:2013gha}.

\subsection{Fermionic dark matter}

As the simplest model inevitably leads to thermalization of the visible and hidden sectors, in the remaining of this paper we consider a model where in addition to the new scalar field $s$ the hidden sector contains a fermion $\psi$ which comprises the dark matter abundance we observe today. Similar model has been studied in the standard DM freeze-out and freeze-in scenarios, in e.g \cite{Kim:2008pp,LopezHonorez:2012kv,Fairbairn:2013uta,Klasen:2013ypa,Alanne:2014bra,Kouvaris:2014uoa,Kainulainen:2015sva,Kainulainen:2016vzv}. 

The Lagrangian of the model is
\be
\mathcal{L} \supset \frac{\mu_{\rm hs}}{2} s h^2 +  \frac{\lambda_{\rm hs}}{4} s^2 h^2 + \frac{\mu_3}{3} s^3 + \frac{\lambda_{\rm s}}{4} s^4 + y s \bar{\psi}\psi.
\ee
If the SM thermalizes with the hidden sector before $\psi$ freezes out, we end up to the standard freeze-out scenario. We neglect the $\lambda_{\rm hs}$ coupling assuming that the $\mu_{\rm hs}$ dominates the SM heating, and take sufficiently small $\mu_{\rm hs}$  to prevent the SM sector from thermalizing with the hidden sector. Further, we assume $m_{\rm s}<2m_\psi$ so that $s$ decays only to SM particles. Decays to SM particles occur via mixing of $s$ and $h$, and if $m_{\rm s}>2m_{\rm h}$ $s$ can also decay directly to Higgs bosons. The mixing angle below $T_{\rm SM}\sim 150$\,GeV is
\be \label{mixingangle}
\tan(2\beta) \simeq \frac{2 v\mu_{\rm hs}}{m_{\rm h}^2-m_{\rm s}^2},
\ee
where $v=246$\,GeV is the vacuum expectation value of the Higgs field. The decay width of $s$ is given by
\be \label{sdecay}
\Gamma_s = \frac{\mu_{\rm hs}^2}{8\pi m_{\rm s}} \sqrt{1-\frac{4m_{\rm h}^2}{m_{\rm s}^2}} + \sin^2(\beta) \Gamma_h(m_{\rm s}),
\ee
where $ \Gamma_h(m_{\rm s})$ is the off-shell Higgs boson decay width, which we evaluate similar to Ref. \cite{Cline:2013gha}.

The $\psi$ abundance freezes out when the annihilation rate $n_\psi\left\langle v\sigma_{\psi\bar\psi\to ss}\right\rangle$ becomes smaller than the Hubble rate $H$. The annihilation cross section in the non-relativistic limit for $m_{\rm s}^2\ll m_{\psi}^2$ is given by
\be
\left\langle v\sigma_{\psi\bar\psi\to ss}\right\rangle \simeq \frac{3 y^2 T_{\rm h} (\mu_3 + 2 m_\psi y)^2}{128\pi m_\psi^5}.
\ee
After the $\psi$ freeze-out, $s$ may still remain in thermal equilibrium with itself until the rates of processes $sss\to ss$ and $ssss\to ss$ drop below $H$. Non-relativistic cross sections for these processes are given by
\be
\label{3to2}
\left\langle v^2\sigma_{sss\to ss}\right\rangle = \frac{25 \sqrt{5} \mu_3^2 (9 \lambda_{\rm s} m_{\rm s}^2 - 2 \mu_3^2)^2}{1536 \pi m_{\rm s}^{11}},
\ee 
and
\be
\label{4to2}
\left\langle v^3\sigma_{ssss\to ss}\right\rangle = \frac{\sqrt{3} \lambda_{\rm s}^2 (9 \lambda_{\rm s} m_{\rm s}^2 - 2 \mu_3^2)^2}{32 \pi m_{\rm s}^{12}}.
\ee 
In the limit $m_\psi^2\gg m_{\rm s}^2$ other number-changing processes such as $\psi\bar\psi\to ss$ $ss\psi\to s\psi$ and $s\psi\psi\to\psi\psi$ can be neglected because the number density of $\psi$ is very small compared to the number density of $s$ when the latter freezes out. In numerical calculations we check that the processes $sss\to ss$ and $ssss\to ss$ indeed dominate over other processes, and that the freeze-outs occur in the non-relativistic region.

The SM sector becomes dominant when $\Gamma_s\sim H$, which may occur before or after the $s$ freeze-out, depending on the strength of the portal coupling and the rate of processes which hold $s$ in thermal equilibrium. Here we concentrate only on the latter case and comment on the former in Sec. \ref{resultsSection}.

Given the relevant annihilation and decay rates, we numerically solve the evolution of the SM energy density $\rho_\gamma$, and number densities of $s$ and $\psi$ from the Boltzmann equations 
\bea
\label{Boltzmann}
&\dot\rho_\gamma + 4H\rho_\gamma = \Gamma_{s} \rho_{\rm s} \,, \\
&\begin{split} 
\dot{n}_{\rm s} + 3Hn_{\rm s} =& - \Gamma_{s} n_{\rm s} \\ 
&- \tfrac{1}{3!} \left\langle v^2\sigma_{sss\to ss}\right\rangle (n_{\rm s}^3 - n_{\rm s}^2 n_{\rm s}^{\rm eq}) \\ 
&- \tfrac{1}{4!} \left\langle v^3\sigma_{ssss\to ss}\right\rangle (n_{\rm s}^4 - n_{\rm s}^2 (n_{\rm s}^{\rm eq})^2) \,,
\end{split} \\
&\dot{n}_\psi + 3Hn_\psi = -\left\langle v\sigma_{\psi\bar\psi\to ss}\right\rangle (n_\psi^2 - (n_\psi^{\rm eq})^2) \,,
\eea
as a function of the scale factor $a$ which we again normalize such that $a=1$ when $T_{\rm h}=m_{\rm s}$. Similar to Sec. \ref{scalarcase}, we use entropy conservation in the $s$ bath to express the hidden sector temperature $T_{\rm h}$ as a function of the scale factor $a$. This approximation is valid if $\Gamma_s \ll H$ during the $s$ freeze-out.

First, we solve only the Boltzmann equation for $\rho_\gamma$ until $s$ becomes non-relativistic assuming that $s$ and $\psi$ are in thermal equilibrium. Here we use
\be
\label{Hubble}
H = \frac{\dot{a}}{a} = \sqrt{\frac{8\pi}{3} \frac{\rho_\gamma + \rho_{\rm s} + \rho_\psi}{M_{\rm P}^2}} .
\ee 
Then, we neglect the $\psi$ energy density in $H$ and solve simultaneously the Boltzmann equations for $\rho_\gamma$ and $n_{\rm s}$ using non-relativistic approximation for the $s$ energy density in thermal equilibrium, $\rho_{\rm s}^{\rm eq} = (m_{\rm s} + 3T_{\rm h}/2) n_{\rm s}^{\rm eq}$. Finally, we solve the evolution of $n_\psi$ using the results obtained for $\rho_\gamma$ and $n_{\rm s}$. This treatment is justified because after $\psi$ has become non-relativistic its energy density is negligible compared to energy densities $\rho_{\rm s}$ and $\rho_\gamma$. Note that this does not mean that $\psi$ freeze-out could not occur before $s$ becomes non-relativistic. 

\begin{figure}
\begin{center}
\includegraphics[width = 0.46\textwidth]{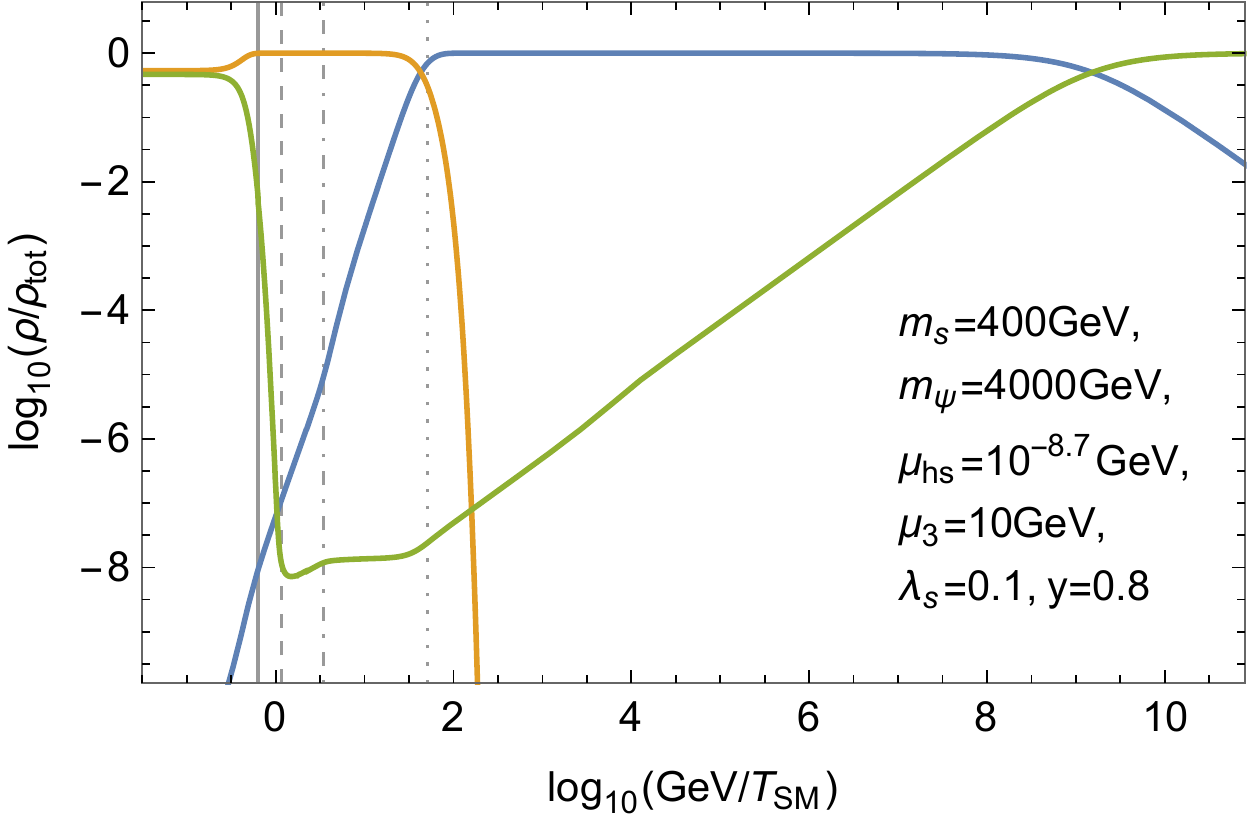} \vspace{0.5cm} \\
\includegraphics[width = 0.46\textwidth]{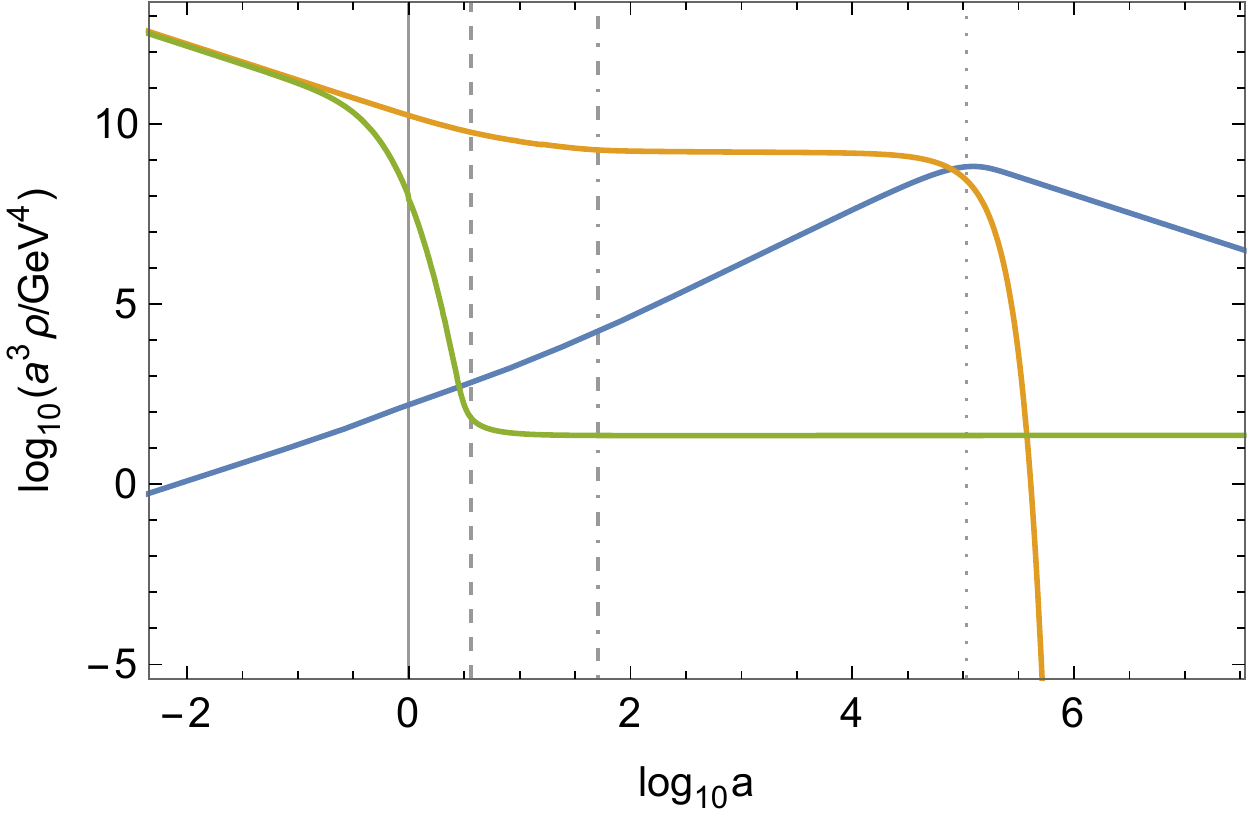}
\caption{Relative energy densities as a function of the SM temperature, and scaling of different components as a function of the scale factor for parameter values shown in the upper panel. Blue, yellow and green lines show the evolution of the SM, $s$, and $\psi$ energy densities, respectively. The solid line shows when $T_{\rm h}=m_{\rm s}$, the dashed line marks the DM freeze-out at $n_\psi \left\langle v\sigma_{\psi\bar\psi\to ss}\right\rangle=3H$, the dot-dashed line marks the $s$ freeze-out at $\tfrac{1}{3!}n_{\rm s}^2\left\langle v^2\sigma_{sss\to ss}\right\rangle=3H$, and the dotted line shows when $\Gamma_s=3H$.}
\label{abundances}
\end{center}
\end{figure}

Fig. \ref{abundances} depicts generic features of evolution of the hidden sector population and the SM heating for parameters which give the correct DM abundance and for which the SM dominates the evolution of the Universe at the time of BBN at $T_{\rm SM}\simeq 4$\,MeV. The lower panel shows the scaling of different components. In the region left from the solid vertical line where $s$ is relativistic the SM energy density scales as $\rho_\gamma\sim a^{-2}$, and after the $s$ freeze-out as $\rho_\gamma\sim a^{-3/2}$ until the decay of $s$. After this, the SM energy density dominates the Universe and scales down as radiation, $\rho_\gamma\sim a^{-4}$. The SM scaling can be related to scaling of $s$ energy density, by writing the Boltzmann equation \eqref{Boltzmann} for the SM energy density as
\be
\frac{H}{a^3}\frac{\rm d}{{\rm d}a} \left( a^4\rho_\gamma \right) = \Gamma_s \rho_{\rm s}.
\ee
We see that the SM energy density scales as $\rho_\gamma\sim a^{-k/2}$, if $s$ particles dominate the Universe and their energy density scales as $\rho_{\rm s}\sim a^{-k}$.

The SM temperature at the time the SM sector finally becomes the dominant energy density component can be estimated analytically. The SM energy density produced via $s$ decays is given by
\be
\rho_\gamma \simeq \rho_{\rm s} \frac{\Gamma_s}{4H}.
\ee
Hence, the moment $\rho_\gamma = \rho_{\rm s}$ occurs when
\be \label{smseq}
\frac{\Gamma_s}{4} = H = \sqrt{\frac{8\pi}{3} \frac{2\rho_\gamma}{M_{\rm P}^2}}.
\ee
Combining then Eqs. \eqref{rhogamma} and \eqref{smseq}, the resulting SM temperature can be solved. For the parameters used in Fig. \ref{abundances}, the temperature at $\rho_\gamma = \rho_{\rm s}$ is $T_{\rm SM} \approx 16$\,MeV, which is in reasonable agreement with the numerical result $T_{\rm SM} \approx 19$\,MeV.

\section{Isocurvature perturbations}
\label{isocSection}

Next, we confront our scenario against observational limits on isocurvature perturbations constituted by a primordial Higgs condensate. The existence of such a condensate is expected, as scalar fields which are light, $d^2V(\phi)/d\phi^2 < H^2_*$, and energetically subdominant, $\rho_\phi \ll \rho_{\rm inf}$, during cosmic inflation typically acquire large fluctuations proportional to the inflationary scale $H_*$ \cite{Starobinsky:1994bd}. This is the case especially for the SM Higgs field \cite{Enqvist:2013kaa, Kusenko:2014lra, Figueroa:2015rqa} (for a possible caveat considering the Higgs vacuum instability or a large non-minimal coupling to gravity, see \cite{EliasMiro:2011aa,Kobakhidze:2013tn,Fairbairn:2014zia,Hook:2014uia,Espinosa:2015qea} and \cite{Herranen:2014cua,Herranen:2015ima}, respectively).

Assuming vacuum stability up to $H_*$ and absence of a non-minimal coupling to gravity, the resulting displacement of the Higgs field from its vacuum creates an effective condensate with a typical field value $h_{*}\equiv \sqrt{\langle h^2\rangle} \simeq 0.363 H_*\lambda_{\rm h}^{\scriptscriptstyle -1/4}$ \cite{Enqvist:2013kaa}, where $\lambda_{\rm h}$ is the Higgs boson quartic self-coupling. On top of the homogeneous condensate the Higgs field acquires perturbations which are a priori uncorrelated with perturbations in the hidden sector. Therefore, the Higgs generates an isocurvature perturbation between cold dark matter and radiation,
\be
\label{isocurvature}
S=\frac{\delta\rho_{c}}{\rho_{c}}-\frac{3}{4}\frac{\delta\rho_{\gamma}}{\rho_{\gamma}} ,
\ee
strongly constrained by observations of the Planck satellite. 

To utilize these constraints, we divide the energy density of the baryon-photon fluid, $\rho_\gamma$, into a part which was sourced by the Higgs condensate, $\rho_{\gamma}^{h_0}$, and to a part which was sourced by the hidden sector, $\rho_{\gamma}^{H}$, and which inherited its perturbation spectrum from the hidden sector, $\delta\rho_c/\rho_c = (3/4)\delta\rho_{\gamma}^{H}/\rho_{\gamma}^{H} \simeq -3 \zeta$. Here $\zeta$  is the curvature perturbation. The spectrum of isocurvature perturbations, $\mathcal{P}_{S} = \langle S^2 \rangle \mathcal{P}_{\zeta}/\langle \zeta^2 \rangle$, can thus be written as
\bea
\mathcal{P}_{S} &= \left(\frac{\rho_{\gamma}^{h_0}}{\rho_{\gamma}^{h_0}+\rho_{\gamma}^{H}}\right)^2\left(9+ \left(\frac{3}{4}\right)^2\frac{\mathcal{P}_{\delta h_0}}{\mathcal{P}_{\zeta}}\right)\mathcal{P}_{\zeta} \\
&\equiv \frac{\beta}{1-\beta}\mathcal{P}_{\zeta} ,
\eea
where we used $\langle \delta\rho^{h_0}_{\gamma} \delta\rho_c\rangle=0$ and denoted $\delta h_0=\delta\rho_{h_0}/\rho_{h_0}$. Using then $\mathcal{P}_{\delta h_0} = (9/4)(H_*/(2\pi))^2/h_*^2 \simeq 0.4\lambda_{\rm h}^{\scriptscriptstyle 1/2}$ \cite{Kainulainen:2016vzv}, and the Planck result $\mathcal{P}_{\zeta}\simeq 2.2\times 10^{-9}$ \cite{Ade:2015lrj}, we get
\be
\label{isocurvaturelimit}
\frac{\rho_{\gamma}^{h_0}}{\rho_{\gamma}^{h_0}+\rho_{\gamma}^{H}} = 2.3\times 10^{-5}\sqrt{\frac{\beta}{1-\beta}}\lambda_{\rm h}^{-1/4} ,
\ee
for the ratio between the energy density comprising an isocurvature perturbation and the total energy density in the SM sector at the time of photon decoupling at $T_{\rm dec}\sim 0.3$ eV. The Planck satellite constrains $\beta \lesssim 0.05$ \cite{Ade:2015lrj}, so that the isocurvature perturbations have only a negligible effect on the evolution of adiabatic perturbations.

The post-inflationary evolution of the Higgs condensate has been studied in detail \cite{Enqvist:2013kaa,Enqvist:2014zqa,Enqvist:2014tta,Figueroa:2015rqa,Enqvist:2015sua,Lozanov:2016pac}. The homogeneous Higgs field begins to oscillate about its minimum after it becomes massive at $3\lambda_{\rm h}h_*^2=H^2$, and decays into SM particles after $\mathcal{O}(10)$ oscillations. The produced particles form a heat bath whose energy density scales down as radiation, $a^{-4}$. The evolution of $a$ is governed by the hidden sector until the $s$ particles decay into SM. After this, the SM starts to dominate the total energy density of the Universe, as discussed in Sec. \ref{reheating}.

Even though most of the final SM heat bath consists of modes adiabatic with the hidden sector, the isocurvature perturbation sourced by the Higgs condensate persists. Their ratio at the formation of the final SM heat bath at $a_{\rm dom}$ when the SM sector starts to dominate the energy density of the Universe is given by
\be
\label{isocurvaturecontribution}
\frac{\rho_{\gamma}^{h_0}}{\rho_{\gamma}^{h_0}+\rho_{\gamma}^{H}} \simeq \frac{\frac{\lambda_{\rm h}}{4} h_*^4}{\rho_\gamma(a_{\rm dom})}\left(\frac{a_0}{a_{\rm dom}}\right)^4 ,
\ee
where $a_0\simeq1.6 g_{\rm h}^{1/4} \lambda_{\rm h}^{-1/8} m_{\rm s}/\sqrt{H_* M_{\rm P}}$ denotes the time when the Higgs condensate begins to oscillate. Here $g_{\rm h}$ counts for the number of relativistic degrees of freedom in the hidden sector, and $\rho_\gamma$ is given by \eqref{Boltzmann}. The evolution of $a$ is determined by different components at different times, as given by \eqref{Hubble}. Because after heating of the SM the energy densities of both the adiabatic and the isocurvature perturbation scale similarly, the ratio \eqref{isocurvaturecontribution} holds true also at the time of photon decoupling. While this ratio depends on details of inflaton decay (see e.g. \cite{Berlin:2016vnh,Figueroa:2016dsc}), Eq. \eqref{isocurvaturecontribution} provides an absolute upper bound on the contribution of the primordial Higgs condensate (for a possible caveat considering the inflaton undergoing a kination phase, see e.g. \cite{Figueroa:2016dsc}).

By then combining Eq. \eqref{isocurvaturecontribution} with Eq. \eqref{isocurvaturelimit}, we can express the isocurvature parameter as
\be
\beta \simeq \frac{7.8\times10^{-15} A g_{\rm h}(a_0)^2}{\sqrt{\lambda_{\rm h}}} \left(\frac{H_*}{10^{14}\mathrm{GeV}}\right)^4,
\ee
where the inflationary scale is bounded above by the non-observation of primordial tensor modes to $H_*\lesssim 8\times10^{13}$\,GeV \cite{Ade:2015lrj}, and
\be \label{Aeq}
A = \left(\frac{m_{\rm s}^4}{\rho_\gamma(a_{\rm dom})a_{\rm dom}^4}\right)^2,
\ee
which can be calculated by numerically integrating the Boltzmann equation \eqref{Boltzmann}. Here $\rho_\gamma$ is the total energy density in the SM sector, which we approximate as $\rho_\gamma \approx \rho_\gamma^{\rm H}$, as the energy density from the Higgs condensate is always very small. For the parameters in Fig. \ref{abundances} we get $A=5.5\times10^{-8}$.

\begin{figure}
\begin{center}
\includegraphics[width = 0.464\textwidth]{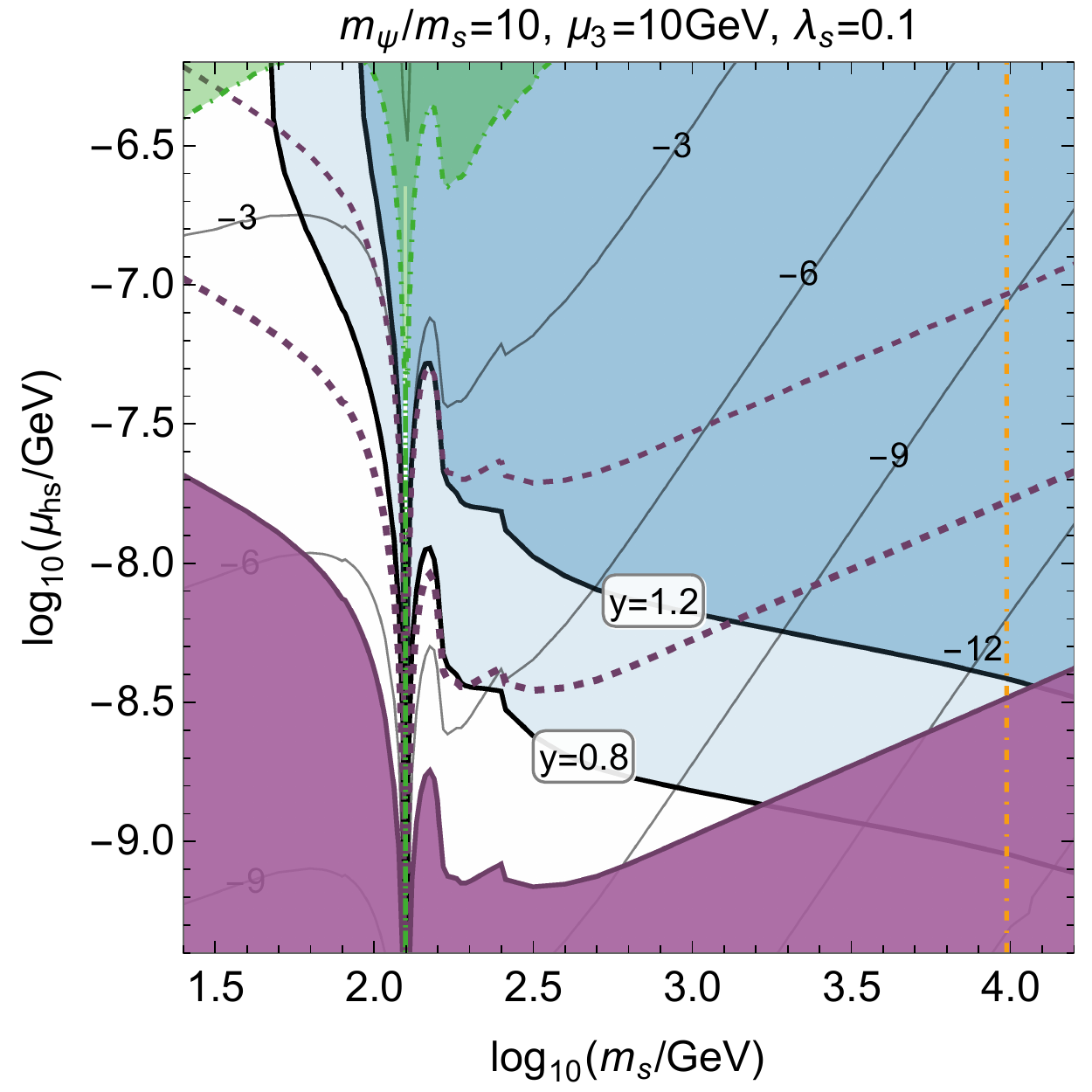} \vspace{0.5cm} \\
\includegraphics[width = 0.464\textwidth]{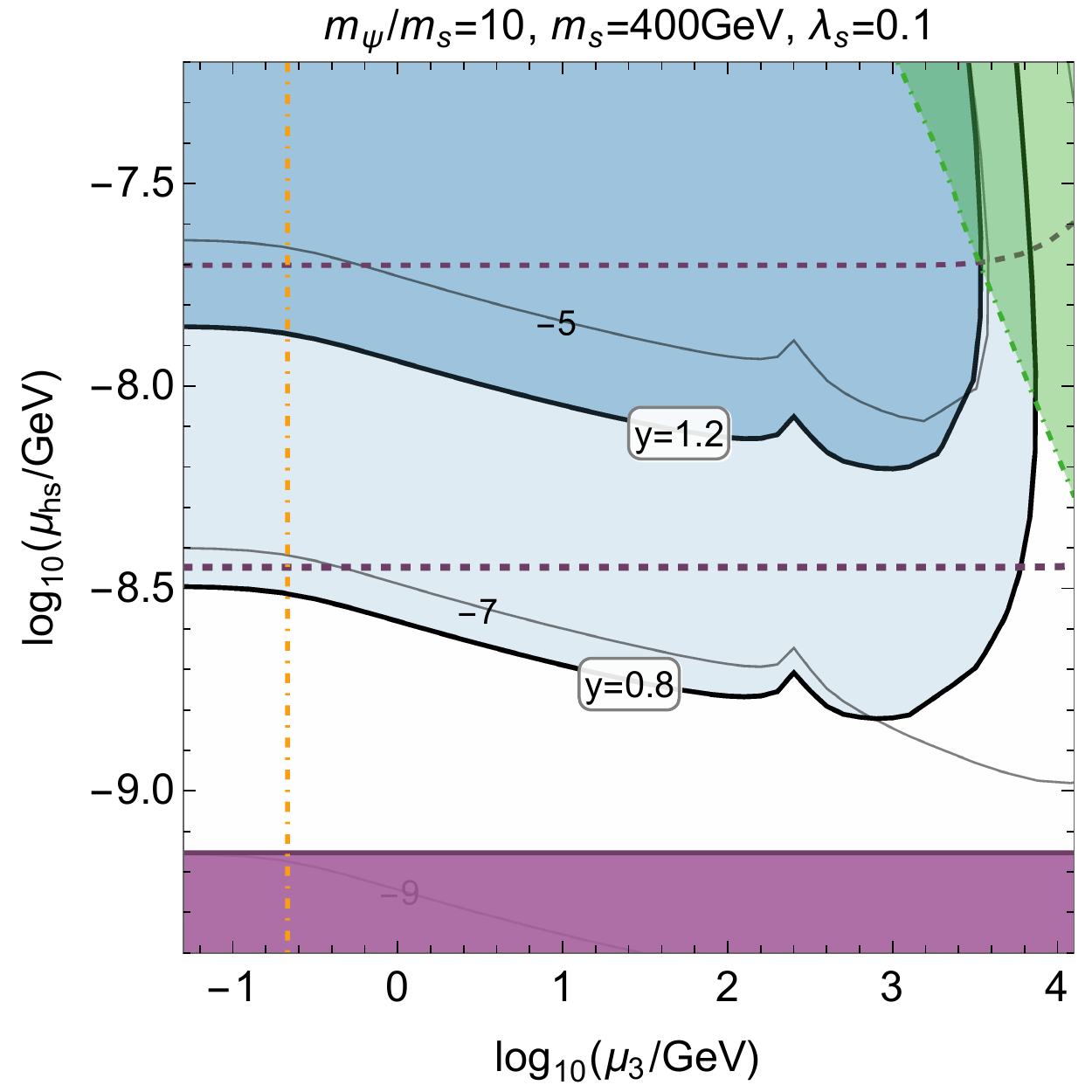} 
\caption{The light and dark blue regions are ruled out by the overproduction of DM for two different values of $y$ shown in the plot. The purple region is excluded by the BBN constraint, $T_{\rm SM}< 4$\,MeV, and the dashed purple lines show where $T_{\rm SM} = 20$\,MeV (thick) and $T_{\rm SM} = 100$\,MeV (thin) at the time the SM sector becomes the dominant energy density component. In the green region $\Gamma_s>3H$ already before the freeze-out of $s$, and there our approximations are not applicable. Left (right) from the vertical yellow dot-dashed line in the lower (upper) panel the $ssss\to ss$ processes determine the $s$ freeze-out instead of $sss \to ss$. Thin contours show $\log_{10}A$, see Eq. \eqref{Aeq}.}
\label{isocurvatureplots}
\end{center}
\end{figure}

\section{Results}
\label{resultsSection}

The dark matter, BBN, and isocurvature bounds for two parameter sets are depicted in Fig. \ref{isocurvatureplots}. We see that the perturbations constituted by a primordial Higgs condensate are typically negligibly small compared to the current upper limit. We also note that because we have neglected the Higgs field's non-minimal coupling to gravity and its possible couplings to other fields during inflation, the bounds shown are strict upper limits on the isocurvature generated in this scenario.

In Fig. \ref{isocurvatureplots} the correct DM abundance is obtained on the black solid line, and successful BBN is retained above the purple region. The moment when the SM sector starts to dominate the Universe is solely determined by the $s$ decay width \eqref{sdecay} via Eq. \eqref{smseq}. The steep change at $m_{\rm s} \sim m_{\rm h}$ in the purple contours, showing the SM temperature at the time the SM energy density starts to dominate the Universe, is caused by cancellation of $m_{\rm s}^2$ and $m_{\rm h}^2$ in the denominator of Eq. \eqref{mixingangle}. The other jumps in the contours as a function of $m_{\rm s}$ are due to the opening of different decay channels.

The shape of the blue region in Fig. \ref{isocurvatureplots}, showing where too large DM abundance is obtained, can be understood as follows: First, if $\mu_3$ is sufficiently small for the $\psi\psi\to ss$ cross section to be determined solely by $t$- and $u$-channel processes, then decreasing $\mu_3$ decreases also the DM abundance relative to radiation as the freeze-out of $s$ occurs earlier. However, in the lower panel left from the vertical yellow line the $s$ freeze-out is determined by $ssss\to ss$ which is (almost) independent of $\mu_3$. Second, if $\mu_3$ is large enough for the $s$-channel diagram to dominate the $\psi\psi\to ss$ scattering, then increasing $\mu_3$ decreases the DM abundance since in that case $\psi$ remains longer in thermal equilibrium before it freezes out. Increasing $\mu_{\rm hs}$ increases DM abundance because it increases the SM temperature when the SM sector becomes dominant, thus extending the radiation dominated era. Decreasing the Yukawa-coupling $y$ increases the region where DM is overproduced, as is standard for the freeze-out mechanism. Finally, steep changes in the isocurvature and DM contours in the lower panel are due to cancellation of terms in Eqs. \eqref{3to2} and \eqref{4to2}. 

In Fig. \ref{isocurvatureplots} we have shown a region where the methods used in solving the Boltzmann equations \eqref{Boltzmann} are applicable. If the $s$ freeze-out occurs when $\Gamma_{\rm s}\ll H$, we can approximate that during the $s$ freeze-out entropy is conserved in the $s$ bath. In the green region in Fig. \ref{isocurvatureplots}, where $\Gamma_{\rm s}\sim H$ already before the $s$ freeze-out, one should consider entropy conservation not only in the $s$ bath but together in the SM and $s$ baths, which is numerically stiff. However, as our purpose has been to illustrate the viability of the model, we leave the detailed investigation of the region where $\Gamma_s\sim H$ already before the freeze-out of $s$ for future work.

\section{Conclusions}
\label{conclusions}

In this work, we have considered cosmological constraints on a scenario where the Standard Model is thermally decoupled from a hidden sector which sources the SM heat bath. Concretely, we considered a hidden sector interacting with the SM fields only through a Higgs portal. 

First, we showed how the simplest scenario, where the hidden sector consists of a $Z_2$-symmetric scalar $s$ only, cannot heat up the SM sector without thermalizing the visible sector with the hidden sector. Then, we extended the model with a singlet fermion $\psi$ and allowed for a general renormalizable scalar potential. We demonstrated how the SM heat bath is generated by scalar decays prior to the Big Bang Nucleosynthesis and how, already prior to this, the $\psi$ abundance freezes out from the hidden sector bath to comprise the observed DM abundance.

We tested our scenario against cosmological constraints and found it is well in line with bounds for the dark matter abundance, BBN, and primordial isocurvature perturbations constituted by a Higgs condensate. For the first time, we computed the isocurvature bound in this scenario and showed it is not problematic for the viability of the model.

However, requiring the SM sector to remain thermally decoupled from the hidden sector causes the SM to acquire a relatively small temperature at the time it becomes the dominant component; in parts of the parameter space the reheating temperature can be considerably low, $T_{\rm SM}\simeq \mathcal{O}(10)$\,MeV. While this is not a problem for retaining a successful BBN, work remains to be done in e.g. considering the viability of models for baryogenesis in these kinds of scenarios. Earlier studies have, however, shown that baryogenesis may be much less difficult than expected with a low reheating temperature \cite{Giudice:2000ex,Davidson:2000dw,Allahverdi:2010im}. 

A low reheating temperature may also allow scenarios where the amplitude of dark matter density perturbations becomes enhanced, for example by virtue of an early matter dominated era before SM reheating. This can lead to observable deviations from the standard predictions for the abundance of Earth-mass or smaller dark matter microhalos \cite{Erickcek:2011us}. Because in this work we concentrated on a scenario where the SM sector becomes dominant only after the $s$ freeze-out, there indeed is an era of matter dominance before SM reheating, as shown in the lower panel of Fig. \ref{abundances}. Moreover, in this kind of a scenario the dark matter freeze-out can occur much earlier than in the standard WIMP scenario, which also enhances the amplitude of dark matter density perturbations.

Hence, it would be interesting to further investigate scenarios where hidden sector dynamics lead to a non-trivial thermal history of the Universe. Probing different baryogenesis scenarios and effects of low reheating temperatures on structure formation would be of particular interest.  Also, one can realize a first order phase transition in the hidden sector, which could produce an observable gravitational wave signal, as has recently been studied in e.g. \cite{Schwaller:2015tja,Jaeckel:2016jlh,Barenboim:2016mjm}. For example, it could be that decay of the hidden sector scalar field to SM particles was possible only after it obtained a non-zero vacuum expectation value. These prospects demonstrate how detailed studies on dynamics of new physics in the early Universe and its imprints on cosmological and astrophysical observables can provide a valuable resource in testing different SM extensions --even in the case where their observable signatures are out of reach of current or near-future direct detection experiments or particle colliders.

\section*{Acknowledgements}
We thank M. Heikinheimo, S. Nurmi, K. Tuominen, and H. Veerm\"ae for discussions. T.T. acknowledges financial support from the Research Foundation of the University of Helsinki and V.V. from the Magnus Ehrnrooth Foundation.

\bibliography{RH.bib}

\end{document}